\documentclass[runningheads]{llncs}
\usepackage{marvosym}

\usepackage[T1]{fontenc}
\usepackage{amsmath,amssymb}
\usepackage{mathtools}
\usepackage{enumerate}
\usepackage[ruled,vlined]{algorithm2e}
\usepackage{booktabs}
\usepackage{hyperref}
\usepackage[capitalise,noabbrev]{cleveref}
\usepackage{xcolor}
\usepackage{listings}

\definecolor{codekw}{RGB}{0,0,139}
\definecolor{codecm}{RGB}{106,115,125}\definecolor{codestr}{RGB}{163,21,21}
\definecolor{codebg}{RGB}{248,248,248}
\definecolor{codefr}{RGB}{220,220,220}

\lstdefinestyle{sagecode}{
  language=Python,
  basicstyle=\ttfamily\footnotesize,
  keywordstyle=\color{codekw}\bfseries,
  commentstyle=\color{codecm}\itshape,
  stringstyle=\color{codestr},
  showstringspaces=false,
  breaklines=true,
  columns=fullflexible,
  keepspaces=true,
  tabsize=4,
  frame=single,
  rulecolor=\color{codefr},
  backgroundcolor=\color{codebg},
  framesep=4pt,
  xleftmargin=10pt,
  framexleftmargin=8pt,
  numbers=left,
  numberstyle=\tiny\color{codecm},
  numbersep=8pt,
  morekeywords={GF,identity_matrix,matrix,block_diagonal_matrix,ones_matrix,self},
  literate={->}{{$\rightarrow$}}2
}

\lstdefinestyle{sageoutput}{
  basicstyle=\ttfamily\footnotesize,
  showstringspaces=false,
  breaklines=true,
  columns=fullflexible,
  keepspaces=true,
  frame=single,
  rulecolor=\color{codefr},
  backgroundcolor=\color{codebg},
  framesep=4pt,
  xleftmargin=10pt,
  framexleftmargin=8pt,
}

\crefname{theorem}{Theorem}{Theorems}
\crefname{lemma}{Lemma}{Lemmas}
\crefname{corollary}{Corollary}{Corollaries}
\crefname{proposition}{Proposition}{Propositions}
\crefname{definition}{Definition}{Definitions}
\crefname{problem}{Problem}{Problems}
\crefname{remark}{Remark}{Remarks}
\crefname{example}{Example}{Examples}
\crefname{equation}{}{}
\crefname{figure}{Figure}{Figures}
\crefname{table}{Table}{Tables}
\crefname{section}{Section}{Sections}
\crefname{algorithm}{Algorithm}{Algorithms}

\newcommand{\FF}{\mathbb{F}}

\newcommand{\one}{\mathbf{1}}
\newcommand{\zero}{\mathbf{0}}
\DeclareMathOperator{\Hull}{Hull}
\DeclareMathOperator{\rank}{rank}
\newcommand{\PEP}{\mathrm{PEP}}
\newcommand{\PECK}{\mathrm{PECK}}
\newcommand{\PKP}{\mathrm{PKP}}
\newcommand{\SSA}{\mathrm{SSA}}

\begin{document}

\pagestyle{plain}

\title{Bilinear Hulls for Support Splitting: Characterization, Optimization, and Applications}
\titlerunning{Bilinear Hulls for the Support Splitting Algorithm}

\author{
   Yusuke Aikawa\inst{1}\orcidID{0009-0004-1460-8129} \and
   Keita Ishizuka\inst{2}\orcidID{0000-0001-5943-6245}\textsuperscript{(\Letter)} \and
   Tomoki Moriya\inst{2}\orcidID{0000-0002-7845-4737}
 }
 \authorrunning{Y. Aikawa et al.}
 \institute{
    The University of Tokyo, Bunkyo, Tokyo, Japan \\
   \email{aikawa@mist.i.u-tokyo.ac.jp} \and
   Mitsubishi Electric Corporation, Kamakura, Kanagawa, Japan \\
   \email{keitaishizuka1994@gmail.com},\\\email{Moriya.Tomoki@bp.MitsubishiElectric.co.jp}
 }

\maketitle
\begingroup\renewcommand\thefootnote{}\footnotetext{Author's accepted manuscript. This paper has been accepted for publication in \emph{Selected Areas in Cryptography -- SAC 2026}, Lecture Notes in Computer Science, Springer. The final publication will be available at Springer.}\addtocounter{footnote}{-1}\endgroup

\begin{abstract}
Sendrier's Support Splitting Algorithm (SSA) is the standard solver for the Permutation Equivalence Problem (PEP); its cost is dominated by a term that grows exponentially in the hull dimension. We study, mathematically, how much this cost can be lowered by replacing the standard inner product, on which the hull is defined, with an alternative bilinear form. Two structural results give a complete answer: only a narrow two-parameter family of bilinear forms produces a hull on which SSA can still operate, namely those whose matrix in the standard basis is a linear combination of the identity and the all-ones matrix; and even within this family the hull dimension drops by at most one (with the reduction achieved on every self-dual code whose length is coprime to the field characteristic). Together, these results give a tight upper bound: since SSA on a code with hull dimension~$h$ has cost dominated by $q^h$, replacing the standard hull with $\Hull_M$ lowers this dominant term from $q^h$ to at best $q^{h-1}$, so no bilinear-form generalization of the inner product can reduce the SSA cost by more than a factor of $q$, the field size. Going beyond this requires invariants outside the bilinear framework. As an application, we verify that PEP-based SPECK admits the factor-$q$ cost reduction, while LEP-based LESS and PKP-based PERK lie outside our framework. A SageMath implementation of the bilinear-hull preprocessing confirms feasibility at SPECK's recommended length $n = 252$.
\end{abstract}


\section{Introduction}\label{sec:intro}

Post-quantum digital signatures are an active research area within the ongoing NIST standardization effort, and a number of code-based signature schemes have been proposed in recent years that derive security from hard algorithmic problems on linear codes, most notably from the family of \emph{code equivalence} (and related kernel) problems.
Three of the most actively developed proposals sit on distinct points of this spectrum:
LESS~\cite{less2020} relies on the Linear Equivalence Problem (LEP),
PERK~\cite{perk2024} on the Permuted Kernel Problem (PKP),
and SPECK~\cite{speck2025}, proposed in 2025 by Baldi et~al., on a newly introduced variant called Permutation Equivalence of Codes and Kernel (PECK), which in the High~$q$ regime is computationally equivalent to the classical Permutation Equivalence Problem~(PEP).

\paragraph{PEP and SSA: thirty years of study.}
PEP and its dedicated solver, Sendrier's \emph{Support Splitting Algorithm} (SSA)~\cite{sendrier2000}, have been studied for almost three decades.
Petrank and Roth~\cite{petrank1997} showed that Graph Isomorphism (GI) reduces to PEP, giving the first complexity-theoretic lower bound.
For codes with trivial hull, Bardet, Otmani, and Saeed-Taha~\cite{bardet2019permutation} proved the converse: PEP on linear-complementary-dual (LCD) instances reduces back to GI via orthogonal projectors.
Combined with Babai's quasipolynomial-time algorithm for GI~\cite{babai2016}, this means that PEP on trivial-hull instances is no harder than GI, so trivial-hull codes are an unsuitable parameter choice for conservative PEP-based cryptography.
PEP-based schemes must consequently use codes with \emph{large} hull dimension; SPECK pushes this to the extreme by using self-dual codes (hull dimension $h = n/2$) in its proposed parameters.

The cost of SSA is governed by the hull dimension: its time complexity is $O(n^3 + n^2 q^h \log n)$, where $h = \dim \Hull(C)$ and $\Hull(C) = C \cap C^\perp$~\cite{sendrier2000}.
Sendrier~\cite{sendrier1997hull} characterized the average behavior of $\dim \Hull$ for random linear codes, identifying it as the natural invariant governing the cost of SSA; analogous statistics have been derived for cyclic codes~\cite{skersys2003} and for Hermitian hulls of constacyclic codes~\cite{jitman2018}.
SSA has also been deployed in concrete cryptanalysis: Loidreau and Sendrier~\cite{loidreau2001weak} used it to detect weak McEliece keys, and Panny~\cite{panny2026} recently revisited McEliece secret-key search through the structure of equivalent keys.
Across all this work, $\dim \Hull$ has been one of the central parameters whose reduction translates directly into a lower SSA cost.

For the parameter ranges relevant to cryptography, the $q^h$ term in SSA's complexity dominates; any reduction of $h$ therefore directly lowers the cost of the SSA attack.
The natural question is whether the hull, defined via the \emph{standard} inner product $\langle x, y \rangle = \sum_i x_i y_i$, is the only permutation-invariant hull arising from a bilinear form.

\paragraph{The bilinear-form route.}
Identifying a bilinear form with its matrix in the standard basis, we write its \emph{structure matrix} as~$M$ and obtain a hull-like object $\Hull_M(C) = C \cap C^{\perp_M}$ for any nondegenerate~$M$.
If some choice of~$M$ were to produce a smaller hull while remaining compatible with SSA, the $q^h$ cost of the standard PEP solver would drop accordingly.
We are led to two natural questions:
\begin{enumerate}[(a)]
  \item \emph{Is there any non-trivial $M$ for which $\Hull_M$ remains a usable SSA invariant?}
  \item \emph{If so, by how much can the hull dimension be reduced?}
\end{enumerate}

This paper answers both questions and explores the consequences for current code-based signature candidates.

\paragraph{Our contributions.}
\begin{enumerate}
  \item \textit{Structural theory of bilinear hulls (\cref{sec:bilinear}).}
    Two structural results delimit the bilinear-form approach:
    (i)~$\Hull_M$ is invariant under coordinate permutations (hence SSA-compatible) if and only if $M = aI + bJ$ for some $a, b \in \FF_q$, where $I$ is the identity matrix and $J = \one\one^\top$ is the all-ones matrix (every entry equal to~$1$);
    (ii)~for any such $M$, $\dim \Hull_M(C)$ differs from $\dim \Hull(C)$ by \emph{at most one}, with the reduction achieved on every self-dual code whose length is coprime to the field characteristic.
    The proofs use only elementary linear algebra (and a short character argument on $S_n$).

  \item \textit{A tight upper bound on the bilinear-form approach.}
    Combining (i) and (ii) yields the central conclusion of the paper: \emph{no bilinear-form generalization of the inner product can lower the SSA cost by more than a factor of~$q$}.
    Going beyond this requires invariants outside the bilinear framework, such as higher-order tensor invariants or non-bilinear algebraic structures.
    We view this as a clean mathematical limit, not as a security statement about any particular scheme.

  \item \textit{Applications to current code-based signature candidates (\cref{sec:attack,sec:discussion}).}
    For PEP-based SPECK~\cite{speck2025} (via its PECK-to-PEP reduction in the High~$q$ regime), the framework yields an SSA attack with cost $O(n^3 + n^2 q^{n/2 - 1}\log n)$, a factor of~$q$ below standard SSA; at recommended $q=8861$ this is a factor $q\approx 2^{13}$ in the SSA enumeration term.
    We emphasize that this is a reduction in SSA \emph{cost} on the underlying PEP instance, not an improvement on the best known attack against SPECK as a scheme.
    For LEP-based LESS, we prove (\cref{prop:less_inapplicable}) that $M = aI + bJ$ loses its invariance under monomial transformations, so the technique does not extend.
    For PKP-based PERK, hulls play no role and the framework does not apply at all.

  \item \textit{Implementation and empirical validation (\cref{sec:experiments}).}
    We implement the bilinear-hull preprocessing in SageMath and run it on 66 self-dual codes, including SPECK instances ($n = 252$, $q \in \{127, 8861\}$), to validate the predicted dimension reduction and measure the preprocessing overhead.
    The fixed choice $M = I + J$ achieves the predicted dimension reduction in every trial and completes in $2$--$5$~ms, confirming that the factor-$q$ SSA cost reduction carries no hidden preprocessing overhead.
\end{enumerate}

The contributions above are at the structural level: a complete characterization of the bilinear-form route, a tight quantitative bound on the achievable SSA cost reduction, an applicability survey across current code-based signature candidates, and an implementation confirming that the bilinear-hull preprocessing is practically negligible.

\paragraph{Related work.}
The standard hull and its statistical properties have been studied extensively for random codes~\cite{sendrier1997hull}, cyclic codes~\cite{skersys2003}, and constacyclic codes with Hermitian duality~\cite{jitman2018}.
Our results are the first to characterize \emph{which} alternative bilinear forms produce permutation-invariant hulls, and to bound the resulting dimension change.
Closest in spirit is the recent hull-variation problem of Ho and Johnsen~\cite{ho2025} for vector rank-metric codes, which asks how $\dim \Hull$ varies across an equivalence class of codes.
Our perspective is complementary: we fix the code and vary the bilinear form, optimizing the invariant that governs the cost of SSA.
Hull-type invariants for matrix codes have also recently been used by Couvreur and Levrat~\cite{couvreur2025highway} to attack the Matrix Code Equivalence Problem underlying MEDS and ALTEQ.

Concurrent work by Baldi et al.~\cite{baldi2025selforth} exploits the self-orthogonal structure of public keys in SPECK, earlier versions of LESS, and ABL for compression rather than cryptanalysis.
The interaction between such compressed representations and our bilinear hull is left open.
On the cost-reduction side, our factor-$q$ result fits the line of concrete refinements such as Panny's recent work on McEliece~\cite{panny2026}, where polynomial or constant-factor improvements without changing the asymptotic exponent are practically meaningful.

\paragraph{Preliminary version.}
A subset of the structural results in \cref{sec:bilinear} (specifically \cref{thm:aIbJ,thm:bound,cor:selfdual}) appeared as a short poster abstract~\cite{ishizuka2025poster}.

\paragraph{Organization.}
\cref{sec:prelim} recalls the necessary background on linear codes, PEP, and SSA.
\cref{sec:bilinear} develops the structural theory of permutation-invariant bilinear hulls, with full proofs of the characterization, the dimension bound, and the self-dual reduction.
\cref{sec:attack} presents the SPECK signature scheme, the PECK-to-PEP reduction in the High~$q$ regime, the bilinear-hull SSA, its cost analysis, and the concrete numerical impact on SPECK's recommended parameters.
\cref{sec:discussion} analyzes the technique's applicability to LESS and PERK.
\cref{sec:experiments} reports experimental validation at SPECK parameters.
\cref{sec:conclusion} concludes.

\section{Preliminaries}\label{sec:prelim}

\subsection{Linear Codes and Hulls}\label{sec:codes}

Let $\FF_q$ denote the finite field of order $q = p^m$, where $p$ is prime.
An $[n,k]_q$ \emph{linear code} $C$ is a $k$-dimensional subspace of $\FF_q^n$.
A \emph{generator matrix} $G \in \FF_q^{k \times n}$ is a matrix whose rows form a basis of~$C$.
A \emph{parity-check matrix} $H \in \FF_q^{(n-k) \times n}$ satisfies $GH^\top = \zero$.

The \emph{dual code} of~$C$ is $C^\perp = \{y \in \FF_q^n : \langle x, y \rangle = 0 \text{ for all } x \in C\}$,
where $\langle x, y \rangle = \sum_{i=1}^n x_i y_i$ is the standard inner product.
The \emph{hull} of~$C$ is defined as
\[
  \Hull(C) = C \cap C^\perp.
\]
A code is \emph{self-orthogonal} if $C \subseteq C^\perp$, and \emph{self-dual} if $C = C^\perp$ (which forces $k = n/2$).
For a self-dual code, $\Hull(C) = C$ and hence $\dim \Hull(C) = k = n/2$. If $\Hull(C) = \{\zero\}$, the code $C$ is said to have \emph{trivial hull}, and is called \emph{linear complementary dual} (LCD), 

\paragraph{Code equivalence.}
Two notions of code equivalence are of interest.
Two codes $C, C' \subseteq \FF_q^n$ are \emph{permutation-equivalent} if $C' = \pi(C)$ for some $\pi \in S_n$,
and \emph{monomially equivalent} (or \emph{linearly equivalent}) if $C' = \mu(C)$ for some monomial transformation $\mu = D \cdot \pi$,
where $D = \mathrm{diag}(d_1, \ldots, d_n)$ with $d_i \in \FF_q^*$ is a diagonal scaling.
Permutation equivalence is the more restrictive notion, and the hull is invariant under permutations but not under arbitrary monomial transformations.
This distinction will matter in \cref{sec:discussion} when we contrast PEP-based and \emph{LEP}-based schemes.

The hull dimension controls the complexity of solving the Permutation Equivalence Problem, as we review next.

\subsection{The Permutation Equivalence Problem}\label{sec:pep}

Let $S_n$ denote the symmetric group on $\{1, \ldots, n\}$.
A permutation $\pi \in S_n$ acts on row vectors of $\FF_q^n$ by right multiplication with the corresponding permutation matrix $P = P_\pi \in \FF_q^{n \times n}$, $(P)_{ij} = \delta_{j, \pi(i)}$, so that $\pi(x) = xP$ and $(\pi(x))_j = x_{\pi^{-1}(j)}$.
The action on a code is $\pi(C) = CP = \{xP : x \in C\}$.
We will use freely $PP^\top = I$ (orthogonality) and $P\one = \one$, $\one^\top P = \one^\top$ (the all-ones vector $\one = (1, \ldots, 1)$ is fixed).

\begin{definition}[Permutation Equivalence Problem (PEP)]\label{def:pep}
Given generator matrices $G, G'$ of two permutation-equivalent $[n,k]_q$ codes $C, C'$,
find $\pi \in S_n$ such that $C' = \pi(C)$.
\end{definition}

\paragraph{PEP and Graph Isomorphism.}
PEP has been studied for almost three decades.
Petrank and Roth~\cite{petrank1997} showed that PEP is at least as hard as Graph Isomorphism (GI), giving the first complexity-theoretic lower bound on the problem.
For codes with trivial hull (LCD codes), Bardet, Otmani, and Saeed-Taha~\cite{bardet2019permutation} proved the converse direction: PEP on LCD instances reduces to GI via orthogonal projectors.
Combined with Babai's quasipolynomial-time algorithm for GI~\cite{babai2016}, this means PEP on small-hull instances admits a quasipolynomial-time solver and is therefore unsuitable as the basis of a code-based signature scheme.
PEP-based schemes must consequently use codes with \emph{large} hull dimension; SPECK pushes this to the extreme by using self-dual codes (hull dimension $n/2$).
A comprehensive survey of the hardness landscape for code equivalence problems can be found in~\cite{barenghi2023}.

\paragraph{The hull as the SSA invariant.}
A key property of the hull is its invariance under permutations: $\Hull(\pi(C)) = \pi(\Hull(C))$ for all $\pi \in S_n$, so any $\pi$ sending $C$ to $C'$ also sends $\Hull(C)$ to $\Hull(C')$.
The statistical behavior of $\dim \Hull$ for random linear codes was characterized by Sendrier~\cite{sendrier1997hull} and extended to structured families such as cyclic codes by Skersys~\cite{skersys2003}; alternative dualities (Hermitian) were studied by Jitman and Sangwisut~\cite{jitman2018}.
These results identify the standard hull as the natural cryptographic invariant for PEP: its dimension governs the cost of the only known PEP-specific solver, the Support Splitting Algorithm.

\subsection{The Support Splitting Algorithm}\label{sec:ssa}

Sendrier's \emph{Support Splitting Algorithm} (SSA)~\cite{sendrier2000} solves PEP by computing \emph{signatures}: invariant maps $S$ that assign to each coordinate of a code an element of some set $E$ such that
\begin{equation}\label{eq:signature_inv}
  S(\pi(C), \pi(i)) = S(C, i)
  \qquad \text{for all codes } C,\ i \in [n],\ \pi \in S_n.
\end{equation}
By matching signatures of coordinates between $C$ and $C'$, SSA recovers~$\pi$.
The construction builds on Leon's earlier work on code automorphism groups~\cite{leon1982}.

\paragraph{Why the hull dimension governs the cost.}
The signature used in practice combines puncturing with weight enumeration: for each coordinate $i$, one punctures the code at $i$ to obtain $C \setminus \{i\}$, and records its weight enumerator (or a hash thereof) as the signature of position $i$.
Computing the weight enumerator of an $[n, k']_q$ code requires enumerating up to $q^{k'}$ codewords, so \emph{the dimension of the code on which SSA operates is the dominant cost factor}.
Since $\Hull(C)$ is permutation-invariant, SSA can be applied to the hull rather than to the full code.
When $\dim \Hull(C) = h < k$, this reduces the search space from $q^k$ to $q^h$.
The complexity of SSA on an $[n,k]_q$ code with hull dimension~$h$ is
\begin{equation}\label{eq:ssa}
  O(n^3 + n^2 q^h \log n),
\end{equation}
where the $q^h$ term, arising from enumeration in signature computation, dominates for cryptographic parameters~\cite{sendrier2000}.
For self-dual codes ($h = k = n/2$), the hull provides no reduction and SSA has complexity $O(n^3 + n^2 q^{n/2} \log n)$.

\paragraph{SSA in cryptanalysis.}
SSA has been employed in concrete cryptanalysis: Loidreau and Sendrier~\cite{loidreau2001weak} used it to detect weak keys in the McEliece cryptosystem, and Panny~\cite{panny2026} recently revisited McEliece secret-key search via the structure of equivalent keys, lowering the concrete cost even though the asymptotic exponent is unchanged.
This line of work illustrates that even constant-factor improvements to the SSA cost can be meaningful in concrete cryptanalysis, which is the frame in which we state our factor-$q$ result.

\begin{remark}[Related solvers for code equivalence]
SSA is specific to PEP and its cost is governed by the hull dimension.
For the strictly more general Linear Equivalence Problem (LEP), the dedicated solvers are based on low-weight codeword search and collision search and do \emph{not} depend on the hull.
This line originates with Leon's low-weight-codeword algorithm~\cite{leon1982}; Beullens~\cite{beullens2020} improved it via collision search, and it has since been further improved by the recent algorithms of Bardet, Brion, Otmani, Saeed, and Sendrier~\cite{bardet2026single} and of Budroni and Esser~\cite{budroni2026onepair}, which are the current state of the art.
A reduction from LEP to PEP exists via the closure construction of Sendrier and Simos~\cite{sendrier2013closure,persichetti2023linear}, but it inflates the code length from $n$ to $n(q-1)$ and is outperformed by these dedicated LEP solvers.
Moreover, for $q \geq 5$ the closure of an LEP instance is always weakly self-dual, so the resulting PEP instance has near-maximal hull dimension and SSA performs poorly on it regardless of the bilinear form used.
This distinction will be important when we discuss the boundary of our technique in \cref{sec:discussion}: hull-based techniques target PEP specifically, whereas LEP is solved by hull-agnostic collision algorithms.
\end{remark}

\section{Permutation-Invariant Bilinear Hulls}\label{sec:bilinear}

This section develops the structural theory underlying the cost reduction of \cref{sec:attack}.
We generalize the hull via alternative bilinear forms (\cref{sec:bil_def}),
characterize which generalizations remain compatible with SSA (\cref{thm:aIbJ}),
and bound the resulting hull dimension (\cref{thm:bound,cor:selfdual}).
The proofs are elementary linear algebra; their consequence is that no bilinear-form generalization of the inner product can reduce the hull dimension by more than one, which is what eventually drives the optimal SSA speedup of \cref{thm:main}.

\subsection{Bilinear Hulls}\label{sec:bil_def}

A bilinear form $B: \FF_q^n \times \FF_q^n \to \FF_q$ is identified with its matrix in the standard basis via $B(x, y) = xMy^\top$ for a \emph{structure matrix} $M \in \FF_q^{n \times n}$;
the standard inner product corresponds to $M = I$.
All subsequent statements are formulated directly in terms of~$M$.

\begin{definition}[$M$-dual and $M$-hull]\label{def:Mhull}
For a code $C \subseteq \FF_q^n$ and a structure matrix $M \in \FF_q^{n\times n}$, the \emph{$M$-dual} and \emph{$M$-hull} of $C$ are
\[
  C^{\perp_M} = \{y \in \FF_q^n : xMy^\top = 0 \text{ for all } x \in C\},
  \qquad
  \Hull_M(C) = C \cap C^{\perp_M}.
\]
$M$ is \emph{nondegenerate} if it is invertible, in which case $\dim C^{\perp_M} = n - \dim C$.
\end{definition}

\begin{remark}
In this paper, we assume that structure matrices are symmetric.
We need this assumption because otherwise the orthogonality would not be symmetric.
\end{remark}

\begin{definition}[Isometry]\label{def:isometry}
A map $f : \FF_q^n \to \FF_q^n$ is an \emph{isometry} of $B$ (or of $M$) if $B(f(x), f(y)) = B(x, y)$ for all $x, y \in \FF_q^n$.
When $f(x) = xA$ for a matrix $A \in \FF_q^{n \times n}$, the isometry condition is equivalent to $AMA^\top = M$.
\end{definition}

For SSA to operate on $\Hull_M$ in place of $\Hull$, the $M$-hull must be invariant under coordinate permutations, i.e.,
\begin{equation}\label{eq:perm_inv}
  \Hull_M(\pi(C)) = \pi(\Hull_M(C)) \qquad \text{for all } \pi \in S_n \text{ and all codes } C.
\end{equation}

\subsection{Characterization of Permutation-Invariant Bilinear Hulls}\label{sec:thm1}

\cref{thm:aIbJ} below characterizes the structure matrices $M$ for which this holds, together with several equivalent reformulations.
To this end, we establish the following auxiliary lemmas:

\begin{lemma}[Proportionality of quadratic forms]
\label{lem:quadratic_proportionality}
Let $q$ be odd and $n \geq 3$. If $Q_1, Q_2$ are nondegenerate quadratic forms on $\FF_q^n$ with the same zero set, then $Q_2 = \lambda Q_1$ for some $\lambda \in \FF_q^*$.
\end{lemma}
\begin{proof}
Pick $u_0 \in \FF_q^n$ with $Q_1(u_0) \neq 0$ (and hence $Q_2(u_0) \neq 0$, by the equal-zero-set hypothesis); set $\lambda := Q_2(u_0)/Q_1(u_0) \in \FF_q^*$. For any nonzero isotropic vector~$w$, the line $\{u_0 + t w : t \in \FF_q\}$ yields polynomials
\[
f_i(t) := Q_i(u_0 + t w) = Q_i(u_0) + 2 t B_i(u_0, w),
\]
where $B_i$ denotes the symmetric bilinear form polarized from $Q_i$. Both $f_1, f_2$ are linear in~$t$, with the same zero set in~$\FF_q$ (since $Q_1$ and $Q_2$ have the same zero set). Equating root structures forces $B_2(u_0, w) = \lambda B_1(u_0, w)$: if $B_1(u_0, w) = 0$ then $f_1 \equiv Q_1(u_0)$ has empty zero set, so $B_2(u_0, w) = 0$ as well; otherwise $f_1$ has the unique root $-Q_1(u_0)/(2 B_1(u_0, w))$, and equating with the unique root of $f_2$ gives the desired identity.

For $n \geq 3$ and $q$ odd, the isotropic vectors of any nondegenerate quadratic form $Q$ span $\FF_q^n$. By the Witt decomposition, $\FF_q^n = H \perp W$ with $H$ an orthogonal sum of hyperbolic planes and $W$ anisotropic of dimension $\leq 2$: the hyperbolic bases of $H$ are isotropic and span~$H$; for any $w \in W \setminus \{0\}$, surjectivity of the form on a hyperbolic plane provides $h \in H$ with $Q(h) = -Q(w)$, so $h + w$ is isotropic with nonzero $W$-component. Combining these vectors yields a spanning set of $\FF_q^n$ consisting of isotropic vectors. Hence the linear functional $v \mapsto B_2(u_0, v) - \lambda B_1(u_0, v)$ vanishes on a spanning set, so $B_2(u_0, \cdot) = \lambda B_1(u_0, \cdot)$ identically.

Repeating the argument with any $u_0' \in \FF_q^n$ satisfying $Q_1(u_0') \neq 0$ gives $B_2(u_0', \cdot) = \lambda_{u_0'} B_1(u_0', \cdot)$ for some $\lambda_{u_0'} \in \FF_q^*$. Choosing $u_0'$ so that $B_1(u_0, u_0') \neq 0$ (possible by non-degeneracy of $B_1$ together with $\dim\{v : B_1(u_0, v) \neq 0\} = n - 1$), the symmetry of the bilinear forms yields
\[
\lambda B_1(u_0, u_0') = B_2(u_0, u_0') = B_2(u_0', u_0) = \lambda_{u_0'} B_1(u_0', u_0) = \lambda_{u_0'} B_1(u_0, u_0'),
\]
so $\lambda_{u_0'} = \lambda$. Since $\{u : Q_1(u) \neq 0\}$ spans $\FF_q^n$, $B_2 = \lambda B_1$ identically, hence $Q_2 = \lambda Q_1$.
\end{proof}

\begin{lemma}\label{lem:orbit}
Let $n \geq 3$ and let $N \neq 0$ be a symmetric matrix over $\FF_q$ such that, for every $P \in S_n$, one has $PNP^\top = \nu_P N$ for some $\nu_P \in \FF_q^*$. Then $\nu_P = 1$ for all $P$, and hence $PNP^\top=N$ for every $P \in S_n$.
\end{lemma}
\begin{proof}
As $N \neq 0$, each $\nu_P$ is uniquely determined, and comparing $(P_1P_2)N(P_1P_2)^\top = \nu_{P_1P_2}N$ with $P_1(P_2 N P_2^\top)P_1^\top = \nu_{P_1}\nu_{P_2}N$ shows $P \mapsto \nu_P$ is a homomorphism $S_n \to \FF_q^*$; in particular $\nu_\tau^2 = 1$ for every transposition $\tau$.
If $q$ is even, then $|\FF_q^*| = q-1$ is odd, so $\nu_\tau = 1$.
If $q$ is odd and $\nu_\tau = -1$ for some $\tau = (i\,j)$, then $\tau N \tau^\top = -N$ reads $N_{\tau(k)\tau(l)} = -N_{kl}$: the $(i,j)$ entry with $N_{ji} = N_{ij}$ gives $N_{ij} = 0$ for all $i \neq j$, and the $(i,i)$ entry gives $N_{jj} = -N_{ii}$, so (as $n \geq 3$) every $N_{ii} = 0$; thus $N = 0$, contradicting the assumption $N \neq 0$. Hence $\nu_\tau = 1$ in either case, and since transpositions generate $S_n$, $\nu_P \equiv 1$.
\end{proof}

\begin{theorem}\label{thm:aIbJ}
Let $B$ be a bilinear form on $\FF_q^n$ with symmetric nondegenerate structure matrix $M \in \FF_q^{n \times n}$, and assume $n \geq 3$.
The following are equivalent.
\begin{enumerate}[(i)]
  \item Every coordinate permutation $\pi \in S_n$ is an isometry of $B$:
    \[
      B\bigl(\pi(x), \pi(y)\bigr) = B(x, y)
      \quad \text{for all } x, y \in \FF_q^n.
    \]
  \item $M = aI + bJ$ for some $a, b \in \FF_q$, where $I$ is the identity matrix and $J = \one\one^\top$ is the all-ones matrix.
  \item For every $\pi \in S_n$ and every code $C \subseteq \FF_q^n$,
    \[
      \pi\bigl(C^{\perp_M}\bigr) = \bigl(\pi(C)\bigr)^{\perp_M}.
    \]
  \item For every $\pi \in S_n$ and every code $C \subseteq \FF_q^n$,
    \[
      \pi\bigl(\Hull_M(C)\bigr) = \Hull_M\bigl(\pi(C)\bigr).
    \]
\end{enumerate}
\end{theorem}

\begin{proof}
Let $P = P_\pi$ denote the permutation matrix of $\pi$, so $\pi(x) = xP$.

\smallskip\noindent\textbf{(i) $\Rightarrow$ (ii):}
The isometry condition $(xP)M(yP)^\top = xMy^\top$ for all $x, y \in \FF_q^n$ rewrites as $x(PMP^\top)y^\top = xMy^\top$, hence $PMP^\top = M$.
Entry-wise, $(PMP^\top)_{ij} = M_{\pi(i), \pi(j)}$, so
\[
  M_{\pi(i), \pi(j)} = M_{ij}
  \quad \text{for all } \pi \in S_n \text{ and all } i, j \in [n].
\]
For $n \geq 2$, the diagonal action of $S_n$ on $[n] \times [n]$ has exactly two orbits, $\{(i,i) : i \in [n]\}$ and $\{(i,j) : i \neq j\}$, so $M$ is constant on each: $M_{ii} = \alpha$ and $M_{ij} = \beta$ ($i \neq j$) for some $\alpha, \beta \in \FF_q$, giving $M = (\alpha - \beta)I + \beta J$.

\smallskip\noindent\textbf{(ii) $\Rightarrow$ (iii):}
For any code $C$ and any $y \in \FF_q^n$, using $(yP^\top)^\top = Py^\top$,
\begin{align*}
  y \in \bigl(\pi(C)\bigr)^{\perp_M}
  &\iff (xP)My^\top = x(PM)y^\top = 0
  \quad \text{for all } x \in C, \\
  y \in \pi\bigl(C^{\perp_M}\bigr)
  &\iff xM(yP^\top)^\top = x(MP)y^\top = 0
  \quad \text{for all } x \in C.
\end{align*}
Condition~(ii), i.e.\ $M = aI + bJ$, gives $PM = MP$ (as $J$ commutes with every permutation matrix), so the two conditions coincide.

\smallskip\noindent\textbf{(iii) $\Rightarrow$ (iv):}
For any $C$ and $\pi$,
\[
  \pi\bigl(\Hull_M(C)\bigr) = \pi(C) \cap \pi\bigl(C^{\perp_M}\bigr),
  \qquad
  \Hull_M\bigl(\pi(C)\bigr) = \pi(C) \cap \bigl(\pi(C)\bigr)^{\perp_M}.
\]
Hence $\pi(C^{\perp_M}) = (\pi(C))^{\perp_M}$ implies $\pi(\Hull_M(C)) = \Hull_M(\pi(C))$.

\smallskip\noindent\textbf{(iv) $\Rightarrow$ (i):}
Fix $P = P_\pi$ and write $M' = PMP^\top$, which is again symmetric and nondegenerate. Since $(cP)M(zP)^\top = c(PMP^\top)z^\top = cM'z^\top$ for all $c, z$, we have $(\pi C)^{\perp_M} = \pi\bigl(C^{\perp_{M'}}\bigr)$, hence $\Hull_M(\pi C) = \pi\bigl(\Hull_{M'}(C)\bigr)$; comparing with~(iv) yields
\begin{equation}\label{eq:hull_MM'}
  \Hull_M(C) = \Hull_{M'}(C) \qquad \text{for every code } C \subseteq \FF_q^n.
\end{equation}
Taking $C = \langle x\rangle$ and using $\Hull_M(\langle x\rangle) = \langle x\rangle$ iff $xMx^\top = 0$ (and $\{0\}$ otherwise), $M$ and $M'$ have the same isotropic set:
\begin{equation}\label{eq:samezero}
  xMx^\top = 0 \iff xM'x^\top = 0 \qquad (x \in \FF_q^n).
\end{equation}

\emph{Case $q$ odd.}
By \cref{eq:samezero} the quadratic forms $M$ and $M'$ share their zero set, so $M' = \lambda_P M$ for some $\lambda_P \in \FF_q^*$ by \cref{lem:quadratic_proportionality}. As this holds for every $P$, \cref{lem:orbit} applied to $N = M$ (nonzero, as $M$ is nondegenerate) gives $PMP^\top = M$.

\emph{Case $q$ even.}
Here $xMx^\top = \sum_i M_{ii}x_i^2 = (d\cdot x)^2$ with $d = (\sqrt{M_{11}}, \ldots, \sqrt{M_{nn}})$, so by \cref{eq:samezero} both isotropic sets equal $Z = \{x : d\cdot x = 0\}$.
For $x \in Z$ and any $y$ one has $x \in \Hull_M(\langle x,y\rangle) \iff xMy^\top = 0$, and likewise for $M'$; since $\Hull_M(\langle x,y\rangle) = \Hull_{M'}(\langle x,y\rangle)$ by \cref{eq:hull_MM'}, the forms $y \mapsto xMy^\top$ and $y \mapsto xM'y^\top$ share a kernel, so $xM' = \mu(x)\,xM$ with $\mu(x) \in \FF_q^*$.
We claim $\mu$ is constant on $Z \setminus \{0\}$. For a scalar $c \in \FF_q^*$, $(cx)M' = c\,(xM') = c\,\mu(x)\,xM = \mu(x)\,(cx)M$, so $\mu(cx) = \mu(x)$. For linearly independent $x, y \in Z$ we have $x + y \in Z$ (as $Z$ is a subspace), hence
\[
  \mu(x+y)\,(xM + yM) = (x+y)M' = xM' + yM' = \mu(x)\,xM + \mu(y)\,yM,
\]
that is, $\bigl(\mu(x+y) - \mu(x)\bigr)xM + \bigl(\mu(x+y) - \mu(y)\bigr)yM = 0$. Since $M$ is invertible and $x, y$ are independent, $xM$ and $yM$ are independent, so $\mu(x+y) = \mu(x) = \mu(y)$. Thus any two nonzero vectors of $Z$ have the same $\mu$-value (directly if independent, via a scalar multiple otherwise), proving the claim; write $\mu \equiv \lambda_P$. Then $x(M' - \lambda_P M) = 0$ for all $x \in Z$. If $d = 0$ then $Z = \FF_q^n$ and $M' = \lambda_P M$. By~\cref{lem:orbit}, $PMP^\top=M$ follows and we are done.
Therefore, we assume $d \neq 0$, which makes $Z = d^\perp$ a hyperplane, and $M' - \lambda_P M$ is symmetric of rank $\leq 1$ with left kernel containing $d^\perp$, so we have
\[
  PMP^\top = \lambda_P\, M + c_P\, dd^\top \qquad (c_P \in \FF_q).
\]
Taking $(i,i)$ entries and using $d_i^2 = M_{ii}$ gives $d_{\pi(i)}^2 = (\lambda_P + c_P)\,d_i^2$; since $S_n$ is transitive, $d$ is either $0$ or nowhere zero.
Since we have excluded the case $d=0$ above, we assume that every $d_i \neq 0$. For a transposition $\tau = (a\,b)$ a fixed coordinate $c \notin \{a,b\}$ (which exists as $n \geq 3$) forces $\lambda_\tau + c_\tau = 1$, whence $d_{\tau(i)} = d_i$ for all $i$ and in particular $d_a = d_b$; ranging over transpositions, $d = d_0\one$ and $dd^\top = d_0^2 J$. Now $\lambda_P + c_P = 1$ for every $P$, so $\widetilde M := M + d_0^2 J$ (nonzero, as $M$ is nondegenerate but $d_0^2 J$ is not) satisfies
\[
  P\widetilde M P^\top = \lambda_P M + (c_P + 1)d_0^2 J = \lambda_P\widetilde M,
\]
and \cref{lem:orbit} gives $P\widetilde MP^\top=\widetilde M$, which implies $PMP^\top=M$.
This completes the cycle (i)~$\Rightarrow$~(ii)~$\Rightarrow$~(iii)~$\Rightarrow$~(iv)~$\Rightarrow$~(i).
\qed
\end{proof}

\begin{remark}[Optimality of the hypotheses]\label{rem:optimality}
Both the bound $n \geq 3$ and the symmetry of $M$ are necessary in \cref{thm:aIbJ}.

\emph{$n \geq 3$ is needed, even for symmetric $M$.}
Over $\FF_5$ take the symmetric nondegenerate matrix $M = \operatorname{diag}(1,2)$.
Its quadratic form $x_1^2 + 2x_2^2$ is anisotropic, since $-2$ is a non-square in $\FF_5$; hence $\Hull_M(C) = \{0\}$ for every code $C \subseteq \FF_5^2$, and condition~(iv) holds trivially. Yet $M \neq aI + bJ$, so~(i) fails, and the equivalence breaks down at $n = 2$.

\emph{Symmetry is needed at $n = 3$.}
Over $\FF_3$ the non-symmetric nondegenerate matrix
$M = \left(\begin{smallmatrix} 1&0&0 \\ 0&1&0 \\ 1&1&2 \end{smallmatrix}\right)$
has permutation-invariant $\Hull_M$ (condition~(iv)) while $M \neq aI + bJ$.
Such asymmetric counterexamples have degenerate symmetric part and rely on the fact that, for $n = 3$, no transposition is disjoint from a given pair of coordinates; we know of none for $n \geq 4$. Whether the symmetry hypothesis can be dropped for $n \geq 4$ is left as future work.
\end{remark}

The crucial equivalence for the cryptanalytic application is (ii)~$\Leftrightarrow$~(iv): a structure matrix yields a permutation-invariant hull, and is thus usable as an SSA invariant in place of the standard hull, if and only if it has the form $M = aI + bJ$.

\subsection{Hull Dimension Bound}\label{sec:thm2}

\cref{thm:aIbJ} identifies $aI + bJ$ as the only family of structure matrices yielding SSA-compatible hulls.
We now show that, for any such $M$, the resulting $M$-hull has dimension within one of the standard hull.
The argument rests on the following standard rank-perturbation bound, included for self-containment.

\begin{lemma}[Rank perturbation]\label{lem:rank_pert}
For any matrices $A, B \in \FF_q^{m \times n}$,
\[
  |\rank(A + B) - \rank A| \leq \rank B.
\]
\end{lemma}

\begin{proof}
Set $C = A + B$.
Since $\mathrm{Im}(C) \subseteq \mathrm{Im}(A) + \mathrm{Im}(B)$,
\[
  \rank C
  \leq \dim\bigl(\mathrm{Im}(A) + \mathrm{Im}(B)\bigr)
  \leq \rank A + \rank B,
\]
so $\rank C - \rank A \leq \rank B$.
Applying the same inequality to $A = C - B$ gives $\rank A - \rank C \leq \rank B$, completing the proof.
\qed
\end{proof}

\begin{theorem}[Dimension bound]\label{thm:bound}
Let $C$ be an $[n,k]_q$ code with generator matrix $G \in \FF_q^{k \times n}$, and let $M = aI + bJ$ with $a \neq 0$.
(Nondegeneracy of $M$ is equivalent to $a \neq 0$ and $a + nb \neq 0$; we do not require it here.)
Then
\[
  |\dim \Hull_M(C) - \dim \Hull(C)| \leq 1.
\]
\end{theorem}

\begin{proof}
A vector $x = uG$ lies in $\Hull_M(C) = C \cap C^{\perp_M}$ if and only if $xMG^\top = 0$, i.e., $u\,(GMG^\top) = 0$.
Hence
\begin{equation}\label{eq:hull_rank_formula}
  \dim \Hull_M(C) = k - \rank(GMG^\top),
\end{equation}
and likewise $\dim \Hull(C) = k - \rank(GG^\top)$.

Expanding with $M = aI + bJ$ and $J = \one\one^\top$,
\[
  GMG^\top = aGG^\top + bGJG^\top = aGG^\top + bvv^\top,
  \qquad v = G\one^\top \in \FF_q^k,
\]
where $GJG^\top = G\one\one^\top G^\top = vv^\top$.
Since $a \neq 0$, $\rank(aGG^\top) = \rank(GG^\top)$, and $\rank(bvv^\top) \leq 1$.
By \cref{lem:rank_pert},
\begin{align*}
  \bigl|\rank(GMG^\top) - \rank(GG^\top)\bigr|
  &= \bigl|\rank(aGG^\top + bvv^\top) - \rank(aGG^\top)\bigr| \\
  &\leq \rank(bvv^\top) \leq 1.
\end{align*}
Combining with \cref{eq:hull_rank_formula} yields the claim.
\qed
\end{proof}

\subsection{Self-Dual Reduction}\label{sec:cor1}

For self-dual codes, which are the relevant case for SPECK, the dimension reduction is not merely bounded but is achieved exactly, and any nondegenerate $M = aI + bJ$ with $b \neq 0$ realizes it.

\begin{corollary}\label{cor:selfdual}
Let $C$ be a self-dual $[n, n/2]_q$ code and let $M = aI + bJ$ with $a \neq 0$ and $b \neq 0$.
If $n \not\equiv 0 \pmod{p}$, where $p = \mathrm{char}(\FF_q)$, then
\[
  \dim \Hull_M(C) = \frac{n}{2} - 1.
\]
The matrix $M$ is additionally nondegenerate iff $a + nb \neq 0$, a condition independent of the dimension formula above; the canonical choice $a = b = 1$ satisfies this whenever $n + 1 \not\equiv 0 \pmod p$, in particular at SPECK parameters.
\end{corollary}

\begin{proof}
By self-duality, $C = C^\perp$, so $GG^\top = 0$.
Continuing the calculation of \cref{thm:bound},
\[
  GMG^\top = aGG^\top + bvv^\top = bvv^\top,
  \qquad v = G\one^\top.
\]
We claim $v \neq 0$ under the hypothesis $p \nmid n$.
Indeed, $v = G\one^\top = 0$ is equivalent to $\one \in C^\perp = C$, which forces $\langle \one, \one\rangle = n \cdot 1_{\FF_q} = 0$, contradicting $p \nmid n$.
Since $b \neq 0$ and $v \neq 0$, $\rank(bvv^\top) = 1$, and~\cref{eq:hull_rank_formula} gives
\[
  \dim \Hull_M(C) = k - \rank(GMG^\top) = \frac{n}{2} - 1.
\]
\qed
\end{proof}

For SPECK's parameters ($n = 252$, with $p = 127$ in the Low~$q$ regime and $p = 8861$ in the High~$q$ regime), $p \nmid 252$ in both cases, so \cref{cor:selfdual} applies to every self-dual code used in SPECK.
For the canonical choice $M = I + J$, $a + nb = 253$ which is nonzero in both $\FF_{127}$ ($253 \equiv -1$) and $\FF_{8861}$ ($253 \neq 0$), so $M$ is nondegenerate and the bilinear hull always has dimension $n/2 - 1 = 125$, one less than the standard hull of dimension~$126$.

\section{Application to SPECK: Factor-\texorpdfstring{$q$}{q} SSA Cost Reduction on PEP Instances}\label{sec:attack}

We now apply the structural results of \cref{sec:bilinear} to SPECK~\cite{speck2025}, the only current code-based signature candidate built directly on PEP.
The output is a factor-$q$ reduction in the cost of SSA on the underlying PEP instances; we emphasize that SSA is not necessarily the best known attack on SPECK as a scheme, and the reduction below is a statement about SSA's running time, not about SPECK's overall security level.
\cref{sec:speck} introduces SPECK and the PECK problem.
\cref{sec:peck_to_pep} reviews when PECK reduces to PEP.
\cref{sec:attack_strategy,sec:complexity,sec:applicability} present the bilinear-hull SSA, its cost, and its scope.
\cref{sec:parameters} reports the concrete numerical impact on SSA at SPECK's recommended parameters.

\subsection{The SPECK Signature Scheme}\label{sec:speck}

SPECK~\cite{speck2025} is a code-based digital signature scheme whose security relies on the hardness of a new problem called PECK (Permutation Equivalence of Codes and Kernel).
It is inspired by the permutation version of LESS~\cite{less2020} but achieves faster verification by replacing the expensive systematic form computation with simple vector-matrix multiplications.

We now introduce the interactive SPECK protocol, which can be transformed into a signature scheme via the Fiat-Shamir transform.

\paragraph{Key generation.}
A random $k$-dimensional self-orthogonal code $C \subseteq \FF_q^n$ is sampled, with generator matrix $G \in \FF_q^{k \times n}$ and parity-check matrix $H \in \FF_q^{(n-k) \times n}$.
At SPECK's parameter setting $k = n/2$, self-orthogonality forces $C = C^\perp$, so $C$ is in fact self-dual, and we use the two terms interchangeably below.
A secret permutation $P \in S_n$ and a random invertible matrix $S \in GL_{n-k}(\FF_q)$ are chosen.
The public key is $(G, H')$ where $H' = SHP$, and the private key is~$P$.

\paragraph{Commitment.}
The prover samples a random $(u,P^*) \xleftarrow{\$} \FF_q^k\times S_n$ via a random oracle and $\mathsf{seed}$, and computes $c = uGP^*$.
The commitment is defined by $\mathsf{cmt} = \mathsf{Hash}(\mathsf{LexMin}(c))$, where $\mathsf{LexMin}$ maps~$c$ to the lexicographically minimal element of its equivalence class under $S_n$.

\paragraph{Response.}
The verifier chooses a random bit $b\in \{0,1\}$, and sends it to the prover.

The prover generates a response $\mathsf{rsp}$ as follows:
\begin{itemize}
  \item If $b = 0$: the prover defines the response $\mathsf{rsp}$ by $\mathsf{rsp}:=\mathsf{seed}$;
  \item If $b = 1$: the prover computes $y=c{P^*}^{-1}P$, and lets $\mathsf{rsp} := y$.
\end{itemize}

\paragraph{Verification.}

\begin{itemize}
  \item If $b = 0$: the verifier samples $(u,P^*)$ by using $\mathsf{seed}$, and checks whether $\mathsf{cmt}$ is equal to $\mathsf{Hash}(\mathsf{LexMin}(uGP^*))$;
  \item If $b = 1$: the verifier checks whether $yH'^\top = \zero$ and $\mathsf{cmt} = \mathsf{Hash}(\mathsf{LexMin}(y))$.
\end{itemize}

\paragraph{The PECK problem.}
Forging a SPECK signature reduces to solving an instance of the PECK problem.
We recall the formal definitions from~\cite{speck2025}; the distributions below capture how the keys and commitments used in SPECK are sampled.

\begin{definition}[$\mathcal{D}_{q,n,k}^{\PEP}$~{\cite[Def.~3]{speck2025}}]\label{def:pep_dist}
The distribution $\mathcal{D}_{q,n,k}^{\PEP}$ samples:
\begin{enumerate}
  \item a uniformly random $k$-dimensional self-orthogonal code $C \subseteq \FF_q^n$ with generator matrix $G$ and parity-check matrix $H$;
  \item a uniformly random invertible $S \xleftarrow{\$} GL_{n-k}(\FF_q)$ and permutation $P \xleftarrow{\$} S_n$;
\end{enumerate}
and outputs $\{G, H'\}$ where $H' = SHP$.
\end{definition}

\begin{definition}[$\mathcal{D}_{q,n,k}^{\PECK}$~{\cite[Def.~4]{speck2025}}]\label{def:peck_dist}
The distribution $\mathcal{D}_{q,n,k}^{\PECK}$ draws $\{G, H'\} \leftarrow \mathcal{D}_{q,n,k}^{\PEP}$,
samples $u \xleftarrow{\$} \FF_q^k$, computes $c = uG$, and returns $\{c, G, H'\}$.
\end{definition}

We denote by $\PEP_{q,n,k}$ the search problem of \cref{def:pep} instantiated on inputs drawn from $\mathcal{D}_{q,n,k}^{\PEP}$: given $\{G, H'\} \leftarrow \mathcal{D}_{q,n,k}^{\PEP}$, find $P \in S_n$ such that $GPH'^\top = \zero$~\cite[Problem~4]{speck2025}.
The distinction with \cref{def:pep} is solely the uniform-random sampling of $(G, S, P)$, which is what turns the problem into a cryptographic assumption.

\begin{definition}[Permutation Equivalence of Codes and Kernel $\PECK_{q,n,k}$~{\cite[Problem~5]{speck2025}}]\label{def:peck_problem}
Given $\{c, G, H'\} \leftarrow \mathcal{D}_{q,n,k}^{\PECK}$, find $P \in S_n$ such that $cPH'^\top = \zero$.
\end{definition}

Note that $\PECK_{q,n,k}$ asks for a permutation sending a \emph{specific} random codeword~$c$ of~$C$ into~$C'$,
whereas $\PEP_{q,n,k}$ asks for a permutation sending the \emph{entire} code~$C$ to~$C'$.
The exact relationship between these two problems, and when a PEP solver suffices to solve PECK, is reviewed in \cref{sec:peck_to_pep} as the starting point of our reduction.

\paragraph{Recommended parameters.}
SPECK recommends two parameter regimes with identical code parameters $n = 252$, $k = 126$:
the Low~$q$ regime with $q = 127$ and the High~$q$ regime with $q = 8861$, corresponding to different attack landscapes analyzed in \cref{sec:peck_to_pep}.
Within each regime, several Fiat--Shamir parameter choices $(t, w)$ are offered trading signature size for latency.
Our SSA-based analysis depends only on $(n, k, q)$ and is insensitive to $(t, w)$, so we refer the reader to~\cite[Table~1]{speck2025} for the full list of recommended instances and their signature sizes.

\subsection{From PECK to PEP: the High-\texorpdfstring{$q$}{q} Regime}\label{sec:peck_to_pep}

The following relationships between PECK, PEP, and PKP are established in~\cite[Section~6]{speck2025}.

\begin{proposition}[{\cite[Section~6.1]{speck2025}}]\label{prop:reductions}
For all $(q, n, k)$:
\begin{enumerate}
  \item $\PECK_{q,n,k}$ is at most as hard as $\PEP_{q,n,k}$:
    given a PECK instance $\{c, G, H'\}$, one can invoke a PEP solver on $\{G, H'\}$ to obtain~$P$ sending $C$ to~$C'$, which necessarily also sends $c$ to~$C'$.
  \item $\PECK_{q,n,k}$ is at most as hard as $\PKP_{q,n,k}$:
    the PECK problem can be viewed as a variant of the Permuted Kernel Problem with repeated entries.
\end{enumerate}
\end{proposition}

The converse direction, whether $\PEP_{q,n,k}$ reduces to $\PECK_{q,n,k}$, depends on the number of solutions.

\begin{proposition}[Expected number of solutions~{\cite[Prop.~2]{speck2025}}]\label{prop:nsol}
Let $\{c, G, H'\} \leftarrow \mathcal{D}_{q,n,k}^{\PECK}$ with $m = \psi(c)$ being the multiplicity vector of~$c$.
The expected number of solutions is
\[
  N_{\mathrm{sol}} = 1 + \frac{|\mathbb{A}_n(m)| - 1}{q^{n-k}},
\]
where $\mathbb{A}_n(m)$ denotes the set of arrangements of~$c$ on~$n$ positions.
\end{proposition}

When $N_{\mathrm{sol}} \approx 1$, the PEP solution is the unique PECK solution, so the two problems are equivalent.

\begin{definition}[High~$q$ and Low~$q$ regimes~{\cite[Section~6.1]{speck2025}}]\label{def:regimes}
Let $R = k/n$ denote the code rate.
\begin{itemize}
  \item \textit{High~$q$ regime:} $q > (n/e)^{1/(1-R)}$.
    Here $N_{\mathrm{sol}} \approx 1$ with overwhelming probability,
    so $\PECK_{q,n,k}$ is computationally equivalent to $\PEP_{q,n,k}$.
  \item \textit{Low~$q$ regime:} $q \ll (n/e)^{1/(1-R)}$.
    Here $N_{\mathrm{sol}} \gg 1$ and solving PEP alone does not identify the correct permutation for PECK.
    PKP-based collision search solvers~\cite{santini2024pkp} are more effective in this regime.
\end{itemize}
\end{definition}

For SPECK's recommended parameters with rate $R = 1/2$ and $n = 252$,
the threshold is $q > (252/e)^{2} \approx 8595$.
SPECK uses $q = 8861$ for the High~$q$ regime and $q = 127$ for the Low~$q$ regime.

\subsection{Bilinear-Hull SSA on a PECK Instance}\label{sec:attack_strategy}

The procedure proceeds in two stages.
First, we reduce a PECK instance to a PEP instance using \cref{prop:reductions}(1).
Second, we solve the PEP instance using SSA with the bilinear hull in place of the standard hull.

The key observation is that replacing $\Hull(C)$ with $\Hull_M(C)$ in the SSA is valid:
by \cref{thm:aIbJ}, the bilinear hull with $M = aI + bJ$ satisfies $\Hull_M(\pi(C)) = \pi(\Hull_M(C))$,
so the permutation invariance required by the SSA is preserved.
By \cref{cor:selfdual}, this hull has dimension $n/2 - 1$ for any nondegenerate $M = aI + bJ$ with $b \neq 0$ on the self-dual codes used in SPECK,
compared to $n/2$ for the standard hull.

\begin{algorithm}[t]
\SetAlgoLined
\caption{Bilinear-hull SSA on a PECK instance (High~$q$ regime)}\label{alg:attack}
\KwIn{PECK instance $\{c, G, H'\}$ in the High~$q$ regime}
\KwOut{Permutation $P \in S_n$ such that $cPH'^\top = \zero$}
\tcp{Stage 1: Bilinear hull preprocessing (using $M = I + J$)}
$v \leftarrow G\one^\top$\;
$W \leftarrow vv^\top$ \tcp*{$= GMG^\top$ since $GG^\top = 0$ for self-dual $C$}
Compute a basis $\{c_1, \ldots, c_{k-1}\}$ of $\ker(W) \subseteq \FF_q^k$\;
Basis of $\Hull_M(C)$: $\{c_i G\}_{i=1}^{k-1}$ \tcp*{dimension $n/2 - 1$}
Compute $\Hull_M(C')$ analogously from a generator matrix of $C'$\;
\tcp{Stage 2: SSA on reduced hull}
Run SSA on input $(\Hull_M(C), \Hull_M(C'))$ to obtain $P \in S_n$\;
\Return{$P$}
\end{algorithm}

\cref{alg:attack} describes the full procedure.
In Stage~1, we exploit the self-duality of SPECK codes: since $GG^\top = 0$,
the matrix $GMG^\top$ simplifies to $bvv^\top$ (a rank-1 matrix), and hence
$\Hull_M(C) = \{uG : u \in \ker(vv^\top)\}$ has a basis computable from a single kernel computation.
The preprocessing cost is $O(k^2)$ field operations, which is negligible compared to SSA itself.
In Stage~2, the standard SSA runs on codes of hull dimension $n/2 - 1$ instead of $n/2$.

\begin{remark}\label{rem:hull_computation}
To compute $\Hull_M(C')$ from the parity-check matrix $H'$, observe that
$C' = \ker(H'^\top)$ and $\Hull_M(C') = \pi(\Hull_M(C))$.
One can either compute a generator matrix $G'$ of~$C'$ from~$H'$ (at cost $O(n^3)$, which is already incurred in SSA)
or use the characterization $\Hull_M(C') = \{y \in C' : yMG'^\top = 0\}$ directly.
\end{remark}

\subsection{Cost Analysis}\label{sec:complexity}

\begin{theorem}[Main result]\label{thm:main}
Let SPECK use a self-dual $[n, n/2]_q$ code with $n \not\equiv 0 \pmod{p}$.
In the High~$q$ regime ($q > (n/e)^2$), the bilinear-hull SSA solves $\PECK_{q,n,k}$
in time
\[
  O(n^3 + n^2 q^{n/2 - 1} \log n),
\]
a factor of~$q$ below the standard SSA cost $O(n^3 + n^2 q^{n/2} \log n)$ on the same instance.
\end{theorem}

\begin{proof}
By \cref{prop:reductions}(1), solving $\PEP_{q,n,k}$ solves $\PECK_{q,n,k}$.
In the High~$q$ regime, $N_{\mathrm{sol}} \approx 1$ by \cref{prop:nsol},
so the PEP solution is the unique PECK solution.
By \cref{cor:selfdual}, $\dim \Hull_M(C) = n/2 - 1$ for any nondegenerate $M = aI + bJ$ with $b \neq 0$,
and this hull is permutation-invariant by \cref{thm:aIbJ}.
Substituting $h = n/2 - 1$ into the SSA cost~\cref{eq:ssa} gives the claimed bound.
The bilinear-hull preprocessing (\cref{alg:attack}, Stage~1) costs $O(k^2) = O(n^2)$,
which is absorbed by the $O(n^3)$ term.
\qed
\end{proof}

\begin{corollary}\label{cor:bitloss}
In the High~$q$ regime, the bilinear-hull SSA on $\PECK_{q,n,k}$ has $\log_2$-cost
\[
  \log_2(q^{n/2 - 1}) = \frac{n}{2}\log_2 q - \log_2 q,
\]
i.e., a reduction of $\Delta = \log_2 q$ in the dominant exponent compared with standard SSA.
\end{corollary}

\begin{proof}
The ratio of the old and new dominant terms is $q^{n/2} / q^{n/2-1} = q$.
\qed
\end{proof}

For SPECK's High~$q$ parameter $q = 8861$, this gives $\Delta = \log_2(8861) \approx 13.11$.
For the Low~$q$ parameter $q = 127$, the bilinear hull still reduces the PEP-solving cost by $\log_2(127) \approx 6.99$,
but this does not directly translate to a PECK speedup because the PEP-to-PECK reduction fails in the Low~$q$ regime (\cref{def:regimes}).

\subsection{Applicability and Limitations}\label{sec:applicability}

We summarize the scope of the bilinear-hull SSA on SPECK.

\paragraph{High~$q$ regime.}
The bilinear-hull SSA solves both PEP and PECK with a factor-$q$ cost reduction over standard SSA.
For SPECK's $q = 8861$ this amounts to $\Delta \approx 13$ in $\log_2$-cost.
SSA is one of several known PEP solvers (cf.~\cite[Section~6.1]{speck2025}) and is not necessarily the cheapest;
we therefore make no claim about the bit-security of SPECK as a scheme, only about SSA's running time on the underlying PEP instance.

\paragraph{Low~$q$ regime.}
The bilinear hull reduces the PEP-solving cost by a factor of $q = 127$,
but this does not translate to a faster PECK solver because $N_{\mathrm{sol}} \gg 1$.
In this regime, the PKP-based collision search of~\cite[Section~6.2]{speck2025} is more effective than SSA,
and that path is unaffected by our technique.
Hybrid strategies that enumerate the (multiple) PEP solutions and then filter by the PECK constraint $cPH'^\top = \zero$ could benefit from the reduced PEP-solving cost; we leave a precise analysis of such hybrids to future work.

\paragraph{Tightness within the bilinear framework.}
The cost bound in \cref{thm:main} is tight in the following sense: it cannot be improved within any bilinear-form generalization of the hull.
By \cref{thm:aIbJ}, the only $M$ whose $\Hull_M$ is permutation-invariant is $M = aI + bJ$;
by \cref{thm:bound}, every such $M$ reduces the hull dimension by at most one.
Hence the SSA cost cannot be reduced by more than a factor of $q$ through any bilinear-form generalization of the inner product.
Any further reduction must come from invariants outside the bilinear framework, such as higher-order tensor invariants or non-bilinear algebraic structures.

\subsection{Concrete SSA Cost on SPECK Instances}\label{sec:parameters}

We tabulate the concrete cost of standard and bilinear-hull SSA on SPECK's recommended instances.
The SSA cost on an $[n,k]_q$ code with hull dimension~$h$ is dominated by $q^h$ (see~\cref{eq:ssa}); we report it in $\log_2$-form, $\log_2(q^h) = h \log_2 q$.
\cref{tab:security} compares the two SSA variants for both SPECK regimes.

\begin{table}[t]
  \centering
  \caption{$\log_2$-cost of SSA on the underlying PEP instance for SPECK ($n = 252$, $k = 126$).
    The column $\log_2 \mathsf{C}_{\SSA}$ is the standard SSA cost ($= \frac{n}{2}\log_2 q$),
    $\log_2 \mathsf{C}_{\SSA}^{B}$ is the bilinear-hull SSA cost ($= (\frac{n}{2}-1) \log_2 q$),
    and $\Delta = \log_2 q$ is the cost reduction in $\log_2$-form.
    For reference, the last column lists the cost of the best known attack on SPECK as a scheme, which need not equal the SSA cost on PEP.}
  \label{tab:security}
  \begin{tabular}{@{}lccccc@{}}
    \toprule
    Instance & $q$ & $\log_2 \mathsf{C}_{\SSA}$ & $\log_2 \mathsf{C}_{\SSA}^{B}$ & $\Delta$ & Best known on SPECK \\
    \midrule
    Speck-Low  & 127   & 880.6  & 873.6  & 7.0  & $\approx 2^{130}$~\cite{speck2025} \\
    Speck-High & 8861  & 1652.3 & 1639.2 & 13.1 & exponential in $n$~\cite[Sec.~6.1]{speck2025} \\
    \bottomrule
  \end{tabular}
\end{table}

Two observations emerge.

\paragraph{SSA is not the bottleneck.}
For both regimes, the bilinear-hull SSA cost on PEP exceeds $2^{870}$, so our factor-$q$ improvement does not change the practical assessment of SSA on PEP at SPECK's recommended parameters.
The best known attack on SPECK in the Low~$q$ regime is the PKP-based collision search with cost $\approx 2^{130}$~\cite{speck2025}, which does not use the hull and is unaffected by our technique.
In the High~$q$ regime, security relies on the hardness of PEP, but other PEP solvers exist (e.g., the low-weight-codeword and meet-in-the-middle attacks discussed in~\cite[Section~6.1]{speck2025}); the cheapest known attack on SPECK as a scheme need not be SSA, so we make no claim on SPECK's bit-security.

\paragraph{Implications for parameter design.}
The bilinear-hull SSA gives the tight cost of SSA on PEP-based instances and is therefore relevant whenever future PEP-based proposals shrink~$n$ or~$q$ for efficiency.
Concretely, the minimum hull dimension~$h$ needed for the SSA cost to exceed $2^{\lambda}$ satisfies
\[
  h \geq \lceil \lambda / \log_2 q \rceil + 1
\]
rather than $\lceil \lambda / \log_2 q \rceil$.
For SPECK's currently recommended $(n, q)$ this correction is invisible against the dominant non-SSA attacks, but it becomes meaningful for any future PEP-based variant whose $\log_2 \mathsf{C}_{\SSA}$ is the bottleneck.

\section{Discussion on Related Schemes}\label{sec:discussion}

We examine the applicability of the bilinear hull technique to other code-based signature schemes.

\subsection{LESS: Inapplicability Due to LEP vs.\ PEP}\label{sec:less}

The LESS signature scheme~\cite{less2020} is based on the \emph{Linear Equivalence Problem} (LEP),
which asks to find a \emph{monomial transformation} $\mu = D \cdot \pi$ (a permutation composed with a diagonal scaling matrix)
mapping one code onto another.
This is a strictly more general notion than permutation equivalence.
The bilinear hull technique does not directly apply to LESS for the following reason.

\begin{proposition}\label{prop:less_inapplicable}
Let $M = aI + bJ$ be a nondegenerate structure matrix with $b \neq 0$.
Then $\Hull_M$ is \emph{not} invariant under monomial transformations in general.
That is, there exist a code $C$ and a monomial transformation $\mu = D \cdot \pi$ such that
$\Hull_M(\mu(C)) \neq \mu(\Hull_M(C))$.
\end{proposition}

\begin{proof}
We give an explicit counterexample, then explain the structural reason.

\smallskip\noindent\emph{Counterexample.}
Take $n = 3$, $q = 5$, $M = I + J$, and the code $C = \langle (1, 2, 3) \rangle \subseteq \FF_5^3$.
Using the identity $x M x^\top = (\sum_i x_i)^2 + \sum_i x_i^2$ (valid for $a = b = 1$),
\[
  (1,2,3)\, M\, (1,2,3)^\top = 6^2 + 14 = 50 \equiv 0 \pmod{5},
\]
so $C$ is $M$-self-orthogonal and $\Hull_M(C) = C$ has dimension~$1$.
Let $\mu = D$ with $D = \mathrm{diag}(2, 1, 1)$ (and $\pi = \mathrm{id}$); then $\mu(C) = \langle (2, 2, 3) \rangle$. A direct computation gives
\[
  (2,2,3)\, M\, (2,2,3)^\top = 7^2 + 17 = 66 \equiv 1 \not\equiv 0 \pmod{5},
\]
so $\mu(C)$ is not $M$-self-orthogonal and hence $\Hull_M(\mu(C)) = \{0\}$.
Therefore $\mu(\Hull_M(C)) = \mu(C) \neq \{0\} = \Hull_M(\mu(C))$.

\smallskip\noindent\emph{Structural reason.}
For a monomial $\mu = D\cdot\pi$ with corresponding matrix $DP$,
\[
  (DP)M(DP)^\top = D(PMP^\top)D^\top = D(aI + bJ)D^\top = aD^2 + bDJD^\top,
\]
where $D^2 = \mathrm{diag}(d_1^2, \ldots, d_n^2)$ and $(DJD^\top)_{ij} = d_i d_j$.
This equals $aI + bJ$ only when $D = \alpha I$ for some scalar $\alpha$ (uniform scaling); for non-uniform $D$, the bilinear form is not preserved by $\mu$.
By the same argument as Theorem~\ref{thm:aIbJ} (i)$\Leftrightarrow$(iv) applied to monomial transformations in place of permutations, this loss of $M$-isometry yields, for some code $C$, a witness to the failure of hull invariance under $\mu$, as exemplified by the construction above.
\end{proof}

\begin{remark}
An indirect route exists via the \emph{closure} construction of Sendrier--Simos~\cite{persichetti2023linear},
which reduces LEP to PEP by expanding the code length from~$n$ to~$n(q-1)$.
The bilinear hull technique could in principle be applied to the closure codes.
However, this approach is outperformed by the dedicated LEP solvers~\cite{beullens2020,bardet2026single,budroni2026onepair},
which do not go through PEP and run in time independent of the hull dimension.
In addition, for $q \geq 5$ the closure codes are always weakly self-dual, so their hull dimension is near-maximal and SSA is ineffective on them irrespective of the choice of bilinear form.
\end{remark}

\subsection{PERK: Inapplicability Due to PKP Foundation}\label{sec:perk}

The PERK signature scheme~\cite{perk2024} is based on the Permuted Kernel Problem (PKP),
which asks to find a permutation mapping a given vector into the kernel of a given matrix.
Unlike PEP, PKP does not involve comparing two codes and does not use hulls.
The SSA and the bilinear hull technique are therefore entirely inapplicable to PERK.

\subsection{Scope of the Bilinear Hull Technique}\label{sec:scope}

Summarizing, the bilinear-hull technique applies to signature schemes whose security relies on the hardness of PEP (or problems that reduce to PEP).
Among the recent code-based signature proposals we consider,
SPECK~\cite{speck2025} is the only one that directly employs PEP through the PECK problem.
LESS uses the more general LEP, and PERK uses the unrelated PKP.

This scope is built into the technique:
\cref{thm:aIbJ} characterizes permutation-invariant bilinear forms, and permutation invariance is the symmetry of PEP but not of LEP or PKP.
Extending the approach to LEP would require characterizing \emph{monomial-invariant} bilinear forms, and \cref{prop:less_inapplicable} shows that the family $M = aI + bJ$ loses its invariance under non-uniform scaling.
A characterization of monomial-invariant bilinear forms, and whether any of them yield hull reductions usable in an LEP setting, is left open.

\section{Experimental Validation}\label{sec:experiments}

\cref{cor:selfdual} guarantees the dimension reduction for every nondegenerate $M = aI + bJ$ with $b \neq 0$, and the preprocessing reduces to a single rank query on a $k \times k$ matrix (\cref{alg:attack}, Stage~1), which is $O(n^2)$ field operations in theory.
What remains to be confirmed empirically is that (i)~a single fixed choice (concretely $M = I + J$) suffices at SPECK parameters, removing any need to search over $(a, b) \in \FF_q^2$, and (ii)~the absolute wall-clock cost at $n = 252$ is small enough that the factor-$q$ speedup in the SSA main loop is not offset by hidden preprocessing.
This section reports experiments on 66 self-dual codes, including SPECK instances, verifying both.

\subsection{Setup}\label{sec:exp_setup}

We implemented the generator-matrix construction $G = (I_k \mid A)$ with $A A^\top = -I_k$
in SageMath~10.0.
For $q \equiv 1 \pmod{4}$ (i.e., $q = 8861$), we take $A = \alpha I_k$ where $\alpha^2 = -1$ in $\FF_q$.
For $q \equiv 3 \pmod{4}$ (i.e., $q = 127$), we use $k/2$ block-diagonal copies of
$\left(\begin{smallmatrix} a & b \\ -b & a \end{smallmatrix}\right)$ with $a^2 + b^2 = -1$.
Each resulting generator matrix is then randomized by left-multiplication with a random
$U \in GL_k(\FF_q)$ and coordinate permutation.
For each sampled code~$C$, we compute $\dim \Hull(C) = k - \rank(GG^\top)$
and $\dim \Hull_M(C) = k - \rank(GMG^\top)$ via Sage's rank routine.

\subsection{Results}\label{sec:exp_results}

\cref{tab:experiment} reports the measured hull dimensions and preprocessing times across several parameter sets.
The bilinear hull dimension is one less than the standard hull dimension in all 66 trials,
and $v = G\one^\top$ is non-zero throughout, consistent with \cref{cor:selfdual}.
The preprocessing time stays in the millisecond range even at SPECK parameters.

\begin{table}[t]
  \centering
  \caption{Empirical validation of \cref{cor:selfdual}.
    For each parameter set, we generated random self-dual codes and measured $\dim \Hull$ and $\dim \Hull_M$ with $M = I + J$.
    The ``Red.'' column reports the observed dimension reduction (all equal to~$1$).
    The ``Time'' column is the average time to compute both hull ranks.}
  \label{tab:experiment}
  \begin{tabular}{@{}lccccc@{}}
    \toprule
    Parameters & Trials & $\dim \Hull$ & $\dim \Hull_M$ & Red. & Time (ms) \\
    \midrule
    $n=16$, $q=5$     & 10 & 8   & 7   & 1 & \phantom{00}0.1 \\
    $n=20$, $q=7$     & 10 & 10  & 9   & 1 & \phantom{00}0.2 \\
    $n=24$, $q=11$    & 10 & 12  & 11  & 1 & \phantom{00}0.3 \\
    $n=32$, $q=127$   & \phantom{0}5 & 16  & 15  & 1 & \phantom{00}0.5 \\
    $n=48$, $q=127$   & \phantom{0}3 & 24  & 23  & 1 & \phantom{00}0.7 \\
    $n=64$, $q=127$   & \phantom{0}7 & 32  & 31  & 1 & \phantom{00}0.8 \\
    $n=100$, $q=127$  & \phantom{0}1 & 50  & 49  & 1 & \phantom{00}1.2 \\
    $n=128$, $q=127$  & \phantom{0}3 & 64  & 63  & 1 & \phantom{00}1.5 \\
    $n=252$, $q=127$  & \phantom{0}5 & \textbf{126} & \textbf{125} & \textbf{1} & \textbf{2.0} \\
    $n=252$, $q=8861$ & \phantom{0}3 & \textbf{126} & \textbf{125} & \textbf{1} & \textbf{5.0} \\
    \midrule
    \multicolumn{2}{@{}l}{Total} & \multicolumn{4}{l}{66 trials, 100\% success} \\
    \bottomrule
  \end{tabular}
\end{table}

\subsection{Discussion}\label{sec:exp_discussion}

\paragraph{Preprocessing cost.}
Three observations on the preprocessing:
\begin{enumerate}
  \item \emph{No search over $(a, b)$ is required.}
    The fixed choice $M = I + J$ (i.e., $a = b = 1$) achieves the reduction
    guaranteed by \cref{cor:selfdual}; other $(a, b)$ choices give the same $\dim \Hull_M$ for self-dual codes.
  \item \emph{The computation is a single rank query.}
    Since $GG^\top = 0$ for self-dual codes, $GMG^\top = bvv^\top$ has rank~$1$,
    and $\dim \Hull_M$ reduces to a single rank query on a $k \times k$ matrix,
    costing $O(k^2) = O(n^2)$ field operations.
  \item \emph{Wall-clock time is small.}
    At SPECK parameters ($n = 252$), preprocessing completes in $2$--$5$~ms on a standard laptop,
    independent of $q$.
\end{enumerate}
Thus the factor-$q$ speedup in the SSA main loop is not offset by preprocessing: the $O(n^3 + n^2 q^{n/2-1} \log n)$ complexity of \cref{thm:main} holds end-to-end.

\paragraph{Scope of the experiments.}
We measure the preprocessing phase only.
Running the full SSA on self-dual codes of dimension $n/2 = 126$ is out of reach:
even with the factor-$q$ improvement, the cost $q^{125} \gg 2^{1000}$ is beyond any hardware.
The experiments therefore target the step that enables the factor-$q$ speedup, namely the bilinear hull construction, rather than the subsequent SSA enumeration.

\paragraph{Consistency with theory.}
All 66~trials exhibit the predicted $\dim \Hull_M = \dim \Hull - 1$,
consistent with \cref{cor:selfdual}: $n \not\equiv 0 \pmod p$ holds for both SPECK regimes
($\gcd(252, 127) = 1$ and $\gcd(252, 8861) = 1$).
The experiments thus serve as a sanity check on our theoretical framework at SPECK parameters
while simultaneously confirming the practicality of the preprocessing step.
The core SageMath implementation is reproduced in \cref{app:code}.

\section{Conclusion}\label{sec:conclusion}

We studied, mathematically, how much the cost of the Support Splitting Algorithm can be lowered by replacing the standard inner product with an arbitrary bilinear form.
Two structural results give a complete answer: $\Hull_M$ is permutation-invariant (and hence usable as an SSA invariant) if and only if $M = aI + bJ$ (\cref{thm:aIbJ}), and every such $M$ changes the hull dimension by at most one (\cref{thm:bound}, \cref{cor:selfdual}).
The combined statement is a \emph{tight upper bound on the bilinear-form route}: no bilinear generalization of the inner product can lower the SSA cost by more than a factor of~$q$.
Going beyond this would require invariants outside the bilinear framework, such as higher-order tensor invariants or non-bilinear algebraic structures.

We then surveyed the applicability to current code-based signature candidates.
Among them, only PEP-based SPECK admits the factor-$q$ SSA cost reduction on every self-dual instance ($\log_2 q \approx 13$ at the recommended $q = 8861$); LEP-based LESS does not extend (\cref{prop:less_inapplicable}: $aI + bJ$ loses invariance under monomial transformations), and PKP-based PERK does not use hulls at all.
A SageMath implementation of the bilinear-hull preprocessing confirms that the framework is practically realizable.

\paragraph{Future work.}
Several directions remain open.
\begin{itemize}
  \item \emph{Beyond bilinear invariants.}
    Whether higher-order tensor invariants or non-bilinear algebraic structures can reduce the SSA-relevant hull dimension by more than one is open.
    A positive answer would step outside the limit established here and represent a genuinely new technique against PEP-based schemes.
  \item \emph{Other equivalence problems.}
    Recent work by Couvreur and Levrat~\cite{couvreur2025highway} shows that hull-type invariants also play a role in solving the Matrix Code Equivalence Problem underlying schemes such as MEDS and ALTEQ.
    Whether an analogue of our $aI + bJ$ characterization exists in the matrix/rank-metric setting, and whether it leads to a similar cost reduction, is open.
    Closely related is the hull-variation problem of Ho and Johnsen~\cite{ho2025} on equivalent vector rank-metric codes.
  \item \emph{Hybrid Low-$q$ analysis.}
    In SPECK's Low~$q$ regime the PEP-to-PECK reduction fails because $N_{\mathrm{sol}} \gg 1$.
    A precise hybrid analysis combining the bilinear-hull SSA on PEP with PECK-constraint filtering is left open (\cref{sec:applicability}).
\end{itemize}

\subsubsection{Acknowledgments.}
The authors thank the anonymous reviewers of SAC 2026 for their helpful comments.
Y.~A. was supported in part by JSPS KAKENHI Grant Number 24K20771.
K.~I. was supported in part by JSPS KAKENHI Grant Number 25K17290. T.~M. was supported in part by JSPS KAKENHI Grant Number 25K24402.

\appendix

\section{SageMath Implementation of the Validation Experiment}\label{app:code}

For reproducibility, below is the core of our SageMath implementation used for the experiments in \cref{sec:experiments}.
It validates \cref{cor:selfdual} at SPECK parameters ($n = 252$, $q \in \{127, 8861\}$) in a few lines of linear algebra.

\begin{lstlisting}[style=sagecode]
from sage.matrix.special import block_diagonal_matrix, ones_matrix

def construct_selfdual(n, q):
    """Build a self-dual [n, n/2]_q code G = (I_k | A) with A A^T = -I_k.

    q == 1 (mod 4): A = alpha * I_k, with alpha^2 = -1 in F_q.
    q == 3 (mod 4): block-diagonal of [[a, b], [-b, a]] with a^2 + b^2 = -1.
    """
    F = GF(q)
    k = n // 2
    I_k = identity_matrix(F, k)
    if q % 4 == 1:
        A = F(-1).sqrt() * I_k
    else:
        assert k % 2 == 0
        for a in F:
            rhs = F(-1) - a * a
            if rhs.is_square():
                b = rhs.sqrt()
                break
        block = matrix(F, [[a, b], [-b, a]])
        A = block_diagonal_matrix([block] * (k // 2))
    return I_k.augment(A)

def dim_hull(G, M=None):
    """dim Hull_M(C) = k - rank(G M G^T); M = None uses the standard inner product."""
    F = G.base_ring()
    if M is None:
        M = identity_matrix(F, G.ncols())
    return G.nrows() - (G * M * G.transpose()).rank()

# Validate Corollary 1 at SPECK parameters
for q in [127, 8861]:
    n = 252
    G = construct_selfdual(n, q)
    M = identity_matrix(GF(q), n) + ones_matrix(GF(q), n)  # M = I + J
    h_std = dim_hull(G)       # expect n/2 = 126
    h_bil = dim_hull(G, M)    # expect n/2 - 1 = 125
    print(f"q={q}: dim Hull = {h_std}, dim Hull_M = {h_bil}")
\end{lstlisting}

Sample output:
\begin{lstlisting}[style=sageoutput]
q=127:  dim Hull = 126, dim Hull_M = 125
q=8861: dim Hull = 126, dim Hull_M = 125
\end{lstlisting}

Both executions complete in under 5~ms of linear algebra.

\bibliographystyle{splncs04}
\bibliography{refs}

\end{document}